\documentclass[preprint,preprintnumbers,amsmath,amssymb]{revtex4-1}
\usepackage{everypage}
\usepackage{amsfonts}
\usepackage{amssymb}
\usepackage{latexsym}
\usepackage{amsmath}
\usepackage{eepic}
\usepackage{graphicx}
\usepackage[colorlinks=true]{hyperref}

\newcommand{\be}{\begin{eqnarray}}
\newcommand{\ee}{\end{eqnarray}}

\newcommand{\bo}{\boldsymbol \rho}
\newcommand{\ta}{\boldsymbol \tau}

\begin{document}

\title{Three charges on a plane in a magnetic field: Special trajectories}

\author{M.A.~Escobar-Ruiz,\\[8pt]
Centre de Recherches Math\'ematiques, Universit\'e de Montreal, \\
C.P. 6128, succ. Centre-Ville, Montr\'eal, QC H3C 3J7, Canada\\[8pt]
escobarr@crm.umontreal.ca \\[8pt]
and \\[10pt]
C.A. Escobar \\[8pt]
CENTRA, Departamento de F\'isica, Universidade do Algarve, 8005-139 Faro, Portugal  \\[8pt]
cruiz@ualg.pt
}

\bigskip

\begin{abstract}
As a generalization and extension of JMP {\bf 54} (2013) 022901, the classical dynamics of three non-relativistic Coulomb charges $(e_1, m_1)$, $(e_2, m_2)$ and $(e_3, m_3)$ on the plane placed in a perpendicular constant magnetic field is considered. Special trajectories for which the distances between the charges remain unchanged are presented and their corresponding constants of motion are indicated. For these special trajectories the number of constants of motion is larger than the dimension of the configuration space and hence they can be called \emph{particularly superintegrable}. Three physically relevant cases are analyzed in detail, namely that of three electrons, a neutral system and a {Helium-like} system. The $n$-body case is discussed as well.
\end{abstract}

\maketitle

\newpage

\section{INTRODUCTION}

The three-body problem, the classical motion of three particles with pairwise gravitational potential, has a monumental history capturing the attention of many prominent mathematicians, including Euler, Lagrange, Poincar\'e and Sundman \cite{Poincare}-\cite{W}. This problem is surprisingly difficult to solve, even in the so called restricted three-body problem, corresponding to the simple case of three masses moving in a common plane. The quantum three-body \mbox{problem} is not less important. There exists a large number of states and processes which demand to be treated as three-particle systems and they are standard material in advanced quantum mechanics textbooks \cite{LQM}.

Moreover, under electromagnetic interactions a new physics emerges where the study of two-dimensional three-electron systems in a magnetic field has led to the experimental and theoretical discovery of the anomalous quantum Hall effect \cite{L}.

In a previous work \cite{MAT2013} of one of the authors, to be referred below simply as paper I, the classical motion of two non-relativistic Coulomb charges placed on a plane subject to a perpendicular constant magnetic field was analyzed in detail. In this paper we will discuss the case of three charges. As a by product, this case provides an insight for the planar $N$ body problem in magnetic field which will be addressed in the last section of the present work.

It is well known that the 3D motion of a classical charged particle subject to a constant magnetic field is characterized by four integrals of motion: the Hamiltonian and the three components of the pseudomomentum. Thus, the system is superintegrable. Furthermore, restricted to the transverse plane to the magnetic field direction, the 2D problem, the system becomes maximally superintegrable \cite{McSweenWinternitz} and then all trajectories are closed and periodic \cite{Nekhoroshev}. Needless to say that superintegrabilty plays a fundamental role in the description of exactly solvable models in both classical and quantum mechanics (see \cite{Miller} for a review).

We can ask the natural question: can three charges describe periodic trajectories?. The answer is affirmative although those trajectories are rather distinctive. They appear if certain initial conditions are chosen only, and not for any system of charged particles. This implies that special trajectories indicate the appearance of \emph{particular} \mbox{constants} of motion \cite{Turbiner:2013p}: they are conserved on certain trajectories only. We call these trajectories special or superintegrable. For the two-body case, some examples of these trajectories for two electrons (an integrable system) and two particles with opposite charges (a chaotic system) were found in \cite{Curilef} and \cite{Taut}, respectively.

It may be remarked that for a neutral two-body system in 3D the motion is known to be chaotic. In general, trajectories are not closed (see, e.g., \cite{FriedrichWintgen}). For this reason special trajectories represent a certain order. This behavior remains true for a three-body neutral system on the plane.

The main purpose of this paper is to determine integrals and \emph{particular} constants of motion of three Coulomb charges to classify the initial data associated with special trajectories. It will set up the basis towards the quantum case. Here, we follow the philosophy behind the two body problem where exact analytical solutions of the Schr\"odinger equation were found \cite{AMT}. In \cite{AMT} the key point was to take a \emph{particular} constant of motion derived in paper I \cite{MAT2013}, promoting it to a quantum operator and then seek for its common eigenfunctions with the Hamiltonian.

The structure of the paper is organized as follows: in Section \ref{3eB} we start by looking at the integrals of motion of the Hamiltonian. The formulae of this section can be easily translated to the quantum case. Afterwards, in order to study the issue on separation of variables both the Hamiltonian and the corresponding Newton equations of motion are written in suitable variables, and by doing so the relevant parameters (effective and \emph{coupling} charges) of this problem are revealed. The physical systems to focus in will naturally appear as well.

In Section \ref{ST} we show periodic trajectories for which the relative distance between particles remain unchanged during time evolution. They are not generic, they occur only for specific initial conditions associated with the \emph{particular} constants of motion we are looking for. The three electron, neutral and the Helium-like systems are examples worked out explicitly.

Then, Section \ref{STN} describes the extension of such special trajectories to the general case of $n$ Coulomb charges in a constant magnetic field. Finally, in Section \ref{Concl} we present a summary and discussion of the results.

\section{THREE CHARGES IN A MAGNETIC FIELD}
\label{3eB}

The Hamiltonian which describes three non-relativistic particles $(e_i,\, m_i)$ , $i=1,2,3$, placed on the plane subject to a constant and uniform magnetic field $\mathbf B=B\,\hat {\mathbf {z}}$ perpendicular to the plane has the form
\begin{equation}
\begin{aligned}
{\cal H}  \ = & \ \frac{{({\bf p}_1-e_1\,{\bf A}_{{\bo}_{1}})}^2}{2\,m_1}\ + \ \frac{{({\bf p}_2-e_2\,{\bf A}_{{\bo}_{2}})}^2}{2\,m_2} \  + \ \frac{{({\bf p}_3-e_3\,{\bf A}_{{\bo}_{3}})}^2}{2\,m_3}
\\ & \  +  \  \frac{e_1\,e_2}{|\bo_1-\bo_2|} \ +  \ \frac{e_1\,e_3}{|\bo_1-\bo_3|} \ +  \  \frac{e_2\,e_3}{|\bo_2-\bo_3|}  \ ,
\end{aligned}
\label{Hind}
\end{equation}
where ${\bo}_{i}$ denotes the position vector of particle $i$ and ${\bf p}_i$ its corresponding canonical momentum. We shall henceforth stick to the so called symmetric gauge where the magnetic vector potential is $\mathbf A_{\bf r}=\frac{1}{2}\ (\mathbf B\times \bf r)$. It is well known that the total Pseudomomentum
\begin{equation}
{\bf K} \  \equiv \ (K_x,\,K_y) \ = \ \boldsymbol k_1 + \boldsymbol k_2 + \boldsymbol k_3 \ =\ (\mathbf {p}_1+e_1\,\mathbf A_{{\bo}_{1}}) + (\mathbf {p}_2+e_2\,\mathbf A_{{\bo}_{2}})+ (\mathbf {p}_3+e_3\,\mathbf A_{{\bo}_{3}})\ ,
\label{pseudomomentumIND}
\end{equation}
is an integral of motion \cite{Gorkov}-\cite{Avron}, the Poisson bracket $\{ {\bf K},\,{\cal {H}} \}=0$ vanishes. In (\ref{pseudomomentumIND}), ${\boldsymbol k}_{j}$ is the individual pseudomomentum of particle $j=1,2,3$. The total \emph{canonical} momentum ${\bf L}^{total}_z$
\begin{equation}
{\bf L}^{total}_z \  \equiv \  \boldsymbol \ell_{z_1} + \boldsymbol \ell_{z_2} + \boldsymbol \ell_{z_3} \ =\ ({\bo}_{1} \times {\bf p}_1)\ + \ ({\bo}_{2} \times {\mathbf p}_2) \ + \ ({\bo}_{3} \times {\mathbf p}_3) \ ,
\label{LzT}
\end{equation}
is also conserved, $\{ {\bf L}^{total}_z, {\cal {H}} \}=0$. Hence, the problem is characterized by three integrals of motion $K_{x,y},\ L_z^{total}$, where some $(x,y)$-coordinate system is introduced on the plane. The dimension of the configuration space is six. Therefore, in general the problem (\ref{Hind}) is not integrable, the number of integrals is less than the dimension of the configuration space. The quantities $K_{x,y},\ L_z^{total}$ are not in involution, they obey the algebra
\begin{equation}
\begin{aligned}
&\{ K_x,\,K_y \} \ = \  -Q\,B\ ,
\\ & \{ L^{total}_z ,\,K_x \}\  = \  K_y\ ,
\\ & \{ L^{total}_z ,\,K_y \}\  = \  -K_x\ ,
\label{AlgebraInt}
\end{aligned}
\end{equation}
where
\[
Q\ =\ e_1 \ + \  e_2 \  +  \ e_3 \ ,
\]
is the total (net) charge. The Casimir operator ${\cal C}$ is given by
\begin{equation}
 {\cal C}\ =\ K_x^2 \ + \ K_y^2 \  -  \ 2\,Q\,B\, L^{total}_z \ .
\label{Casimir}
\end{equation}
For the case of a single charge the quantity ${\cal C}$ is, in fact, the Hamiltonian.

Let us introduce Jacobi variables in a standard way
\begin{equation}
\begin{aligned}
&\mathbf R = \mu_1\, \bo_1 + \mu_2\, \bo_2+\mu_3\, \bo_3 \quad \ , \quad \ta_1 =  \bo_2-\bo_1 \quad \ , \quad \ta_2 = \bo_3- (\nu_1\bo_2+\nu_2\,\bo_1)\ , \\ &  \mathbf {P} = {\mathbf p}_1 + {\mathbf p}_2 + {\mathbf p}_3\quad \ , \quad \mathbf p_{\ta_1} = \nu_1\,\mathbf p_{2}-\nu_2\,\mathbf p_{1} \quad \ ,  \quad \mathbf p_{\ta_2} = \mathbf p_{3}(\mu_1+\mu_2)-\mu_3\,(\mathbf p_{1}+\mathbf p_{2}) \ ,
\end{aligned}
\label{CMvar}
\end{equation}
where $\mu_i=\frac{m_i}{M}$ and $\nu_i=\frac{m_i}{m_1+m_2}$ are dimensionless parameters, $M = m_1 + m_2+m_3$ is the total mass of the system. In these coordinates the total Pseudomomentum (\ref{pseudomomentumIND}) becomes
\begin{equation}
{\bf K} \  = \  \mathbf P \ +  Q\,\mathbf A_{\mathbf R} \  - \ e_{c1}\,\mathbf A_{\ta_1} \   -  \  e_{c2}\,\mathbf A_{\ta_2}    \ ,
\label{pseudomomentum}
\end{equation}
where the two coefficients
\begin{equation}
\label{coupchar}
e_{c1}\ = \ \frac{m_1\,m_2}{m_1+m_2}\bigg( \frac{e_1}{m_1} - \frac{e_2}{m_2}   \bigg) \qquad ; \qquad e_{c2} \ = \ \frac{(m_1+m_2)\,m_3}{m_1+m_2+m_3}\bigg( \frac{e_1+e_2}{m_1+m_2} - \frac{e_3}{m_3}   \bigg)   \   ,
\end{equation}
can be called {\it coupling charges}. In general, the Hamiltonian (\ref{Hind}) expressed in Jacobi variables (\ref{CMvar}) does not admit separation of the center of mass (CM) motion. However, for particles with equal Larmor frequency, i.e. $\frac{e_1}{m_1}=\frac{e_2}{m_2}=\frac{e_3}{m_3}=\alpha$, both {coupling charges} vanish $e_{c1}=e_{c2}=0$ and the CM motion do separate from the relative ones. In this case the CM motion is described by the elementary Hamiltonian
\[
{\cal H}_{CM}  \ =  \ \frac{{({\bf P}\,-\,Q\,{\bf A}_{\bf R})}^2}{2\,M} \ ,
\]
which corresponds to that of a particle with charge $Q$ and mass $M$ placed in a constant magnetic field. All CM trajectories are circular and possess the same Larmor frequency of the individual particles, $\frac{Q}{M}=\alpha$.

Now, for convenience we introduce another canonical change of variables
\begin{equation}
\begin{aligned}
&  \  \mathbf {P}^\prime = \mathbf {P} -  e_{c1}\,\mathbf A_{\ta_1} -e_{c2}\,\mathbf A_{\ta_2}
\ \ , \qquad {\mathbf p}_{\ta_1}^\prime = {\mathbf p}_{\ta_1} +  e_{c1}\,\mathbf A_{\mathbf R}\ \ , \qquad {\mathbf p}_{\ta_2}^\prime = {\mathbf p}_{\ta_2} +  e_{c2}\,\mathbf A_{\mathbf R} \ ,
\\ & \ {\mathbf R}^\prime = {\mathbf R} \ \ , \qquad  {\ta_1}^\prime = {\ta_1} \ \ , \qquad  {\ta_2}^\prime = {\ta_2} \ ,
\end{aligned}
\label{CC}
\end{equation}
in which the Hamiltonian (\ref{Hind}), hereafter omitting the primes, takes the form
\begin{equation}
\begin{aligned}
{\cal H}  \ = & \ \frac{  {(   \mathbf{P} -Q\,\mathbf A_{\mathbf R} + 2\,e_{c1}\,\mathbf A_{\ta_1} + 2\,e_{c2}\,\mathbf A_{\ta_2}   )}^2   }{2\,M}
\ + \ \frac{  {( {\mathbf p}_{\ta_1}^2- e_{1eff}\,\mathbf A_{\ta_1}   )}^2   }{2\,\nu_1\,\nu_2\,(m_1+m_2)}
\  +  \  \frac{  {( {\mathbf p}_{\ta_2}^2- e_{2eff}\,\mathbf A_{\ta_2}   )}^2   }{2\,m_3\,(\mu_1+\mu_2)}
\\ & \ +  \     \frac{  e_{c1}^2\,\mathbf A_{\ta_1}^2\,\mu_3 }{2\,(m_1+m_2)} \ + \ \frac{  e_{c2}^2\,\mathbf A_{\ta_2}^2\,\mu_3^2 }{2\,\nu_1\,\nu_2\,(m_1+m_2)} \ +  \ \frac{e_{c1}}{m_1+m_2} \mathbf A_{\ta_1}\,{\mathbf p}_{\ta_2} \  -  \ \frac{e_{c1}\,m_3}{m_1\,m_2}\,\mathbf A_{\ta_2}\,{\mathbf p}_{\ta_1}
\\ &   \   +  \ \frac{e_{c1}}{m_1\,m_2} \,\mathbf A_{\ta_1}\,\mathbf A_{\ta_2}[ e_3\,\mu_1\,\mu_2\,(m_1+m_2) +  e_2\,\mu_1\,\mu_3\,(m_1+m_3) +e_1\,\mu_2\,\mu_3\,(m_2+m_3)   ]
\\ & \ +  \ \frac{e_1\,e_2}{|\ta_1|} \ +  \ \frac{e_1\,e_3}{|\ta_2+\nu_2\,\ta_1|} \ +   \   \frac{e_2\,e_3}{|\ta_2-\nu_1\,\ta_1|}  \ ,
\end{aligned}
\label{HC}
\end{equation}
where
\begin{equation}\label{effcharges}
e_{1eff}= e_2\,\nu_1^2 + e_1\,\nu_2^2 \qquad \ , \qquad e_{2eff}=e_3\,{(\mu_1+\mu_2)}^2+(e_1+e_2)\,\mu_3^2 \ ,
\end{equation}
play the role of two \emph{effective charges}, cf. (\ref{coupchar}). Because of the change of variables (\ref{CC}), the total Pseudomomentum
\[
{\bf K} \ = \ {\bf P} \ + \ Q\,\mathbf A_{\mathbf R} \ ,
\]
coincides with the CM Pseudomomentum. Thus, for a neutral system ($Q=0$) the new CM momentum ${\bf P}$ is conserved, $\{\mathbf {P}, {\cal H} \} = 0$, as one would expect for a free particle \cite{Grotch}. However, separation of CM motion and momentum-like conserved quantities are evidently distinct notions.

The first term in (\ref{HC}) describes the kinetic energy of the CM motion and it can be shown that it is gauge invariant. The second and the third terms correspond to the kinetic energy of two effective charges $e_{1eff}$ and $e_{2eff}$ (\ref{effcharges}), respectively. The Hamiltonian (\ref{HC}) is suitable to study finite mass corrections. In the present work we will focus in the general case rather than limiting cases such that the Born-Oppenheimer approximation where $M \rightarrow \infty$. Also, we do not consider a perturbative approach since an exact treatment is preferable.

For a quantum system of three Coulomb charges, the Hamiltonian $\cal H$, the Pseudomomentum ${\bf K}$ and the angular momentum $L^{total}_z$ can be obtained replacing in the previous formulae the momenta by the corresponding differential operators. In particular, the canonical transformation (\ref{CC}) can be achieved by means of a unitary transformation.

\subsection{Equations of motion}

In this section we switch from the phase space (Hamiltonian) to the configuration space (equations of motion). From (\ref{Hind}) we obtain the Newton equations of motion
\begin{equation}
\begin{aligned}
&m_1\,\ddot{\bo}_1\ =\ e_1\,\dot{\bo}_1\times \mathbf{B} \  +   \  \bigg(\frac{e_1\,e_2}{|\bo_1-\bo_2|^3}(\bo_1-\bo_2) \ + \ \frac{e_1\,e_3}{|\bo_1-\bo_3|^3}(\bo_1-\bo_3)\bigg) \ ,
\\ &
m_2\,\ddot{\bo}_2\ =\ e_2\,\dot{\bo}_2\times \mathbf{B} \  - \  \bigg(\frac{e_1\,e_2}{|\bo_1-\bo_2|^3}(\bo_1-\bo_2)\ - \ \frac{e_2\,e_3}{|\bo_2-\bo_3|^3}(\bo_2-\bo_3)\bigg) \ ,
\\ &
m_3\,\ddot{\bo}_3\ =\ e_3\,\dot{\bo}_3\times \mathbf{B}\  -  \  \bigg(\frac{e_1\,e_3}{|\bo_1-\bo_3|^3}(\bo_1-\bo_3)\ + \ \frac{e_2\,e_3}{|\bo_2-\bo_3|^3}(\bo_2-\bo_3)\bigg) \ ,
\end{aligned}
\label{NEsep}
\end{equation}
($\dot{\bo}\equiv \frac{d}{dt}\bo$). The equations (\ref{NEsep}) are invariant under the changes of parameters
\[
(\,B \rightarrow -B,\,  e_1 \rightarrow -e_1,\,  e_2 \rightarrow -e_2,\, e_3 \rightarrow -e_3\,) \ ,
\]
and reflections
\[
(\,\bo_1 \rightarrow - \bo_1,\,\bo_2 \rightarrow - \bo_2,\,\bo_3 \rightarrow - \bo_3\,) \ .
\]

In Jacobi variables (\ref{CMvar}), the Newton equations (\ref{NEsep}) can be written as
\begin{equation}
\begin{aligned}
& M\ddot{\bf R}\ =\ Q\,\dot{\bf R}\times \mathbf{B} \ - \ {\bf E}_{R}(\dot{\ta_1},\,\dot{\ta_2}) \ ,
\\ &
\tilde m_{1}\, \ddot{\ta}_1\ =\ e_{1eff}\dot{ \ta}_1\times {\bf B} \  -  \  \frac{e_{c1}^2 \,B^2}{2M}\ta_1 \ + \  {\bf E}_{1}({\bf R},\,{\ta}_2,\,\dot{ \ta}_2) \ + \ V_1(\ta_1,\,\ta_2)  \ ,  \\ &
\tilde m_{2}\, \ddot{\ta}_2\ =\ e_{2eff} \,\dot{ \ta}_2\times {\bf B}\  - \ \frac{ e_{c2}^2\, B^2}{2\,M} \ta_2 \ +  \ {\bf E}_{2}({\bf R},\,{\ta}_1,\,\dot{ \ta}_1) \ + \ V_2(\ta_1,\,\ta_2) \ ,
\end{aligned}
\label{Eq}
\end{equation}
where
\begin{equation}
\begin{aligned}
& {\bf E}_{R}(\dot{\ta_1},\,\dot{\ta_2})\ =\   e_{c1}\, \dot{\ta_1}\times \mathbf{B} \  +  \ e_{c2}\, \dot{\ta_2}\times \mathbf{B}\ , \\ &
{\bf E}_{1}({\bf R},\,{\ta}_2,\,\dot{ \ta}_2)\ =\  \frac{e_{c1}\,Q\, B^2}{2\,M}  {\bf R} \ - \  \ \frac{e_{c1}}{M} {\bf K}\times {\bf B} \   -  \  \frac{e_{c1} \,e_{c2}\, B^2}{2\,M} \ta_2 \  +\mu_3\, e_{c1}\, \dot{ \ta}_2\times {\bf B} \ ,  \\ &
{\bf E}_{2}({\bf R},\,{\ta}_1,\,\dot{ \ta}_1)  \ =\  \frac{e_{c2}\,Q\, B^2}{2\,M}{\bf R}\  -  \  \frac{e_{c2}}{M} {\bf K}\times {\bf B}  \   -  \   \frac{e_{c1}\,e_{c2}\, B^2}{2\,M}\ta_1  \  +  \ \mu_3\, e_{c1}\, \dot{ \ta}_1\times {\bf B} \ ,
\end{aligned}
\label{EqeL}
\end{equation}

make sense of electric fields, $\tilde m_{1}=\frac{m_1\,m_2}{m_1+m_2},\, \tilde m_{2}=\frac{(m_1+m_2)\,m_3}{m_1+m_2+m_3}$ and

\begin{equation}
\begin{aligned}
&
V_1(\ta_1,\,\ta_2) \ = \ \frac{e_1\,e_2}{|\ta_1|^3}\, \ta_1 \ -  \  \frac{\nu_1 \,e_2\, e_3}{|\ta_2-\nu_1\,\ta_1 |^3}(\ta_2-\nu_1\,\ta_1)\  +  \ \frac{\nu_2 \,e_1 \,e_3}{|\ta_2+\nu_2\,\ta_1 |^3}(\ta_2+\nu_2\,\ta_1) \ ,
\\ &
V_2(\ta_1,\,\ta_2) \ = \  \frac{ e_2\, e_3}{|\ta_2-\nu_1\ta_1 |^3}(\ta_2-\nu_1\ta_1) \  +  \ \frac{ e_1 \,e_3}{|\ta_2+\nu_2\,\ta_1 |^3}(\ta_2+\nu_2\,\ta_1)  \ ,
\end{aligned}
\label{EqCoul}
\end{equation}

describe Coulomb interactions. In (\ref{EqeL}) the Pseudomomentum

\[
{\bf K}  \ = \ M\,\dot{\bf R}\ - \ Q\,{\bf R}\times \mathbf{B}\ + \ e_{c1}\, {\ta_1}\times \mathbf{B} \ +  \  e_{c2}\, {\ta_2}\times \mathbf{B}  \ ,
\]

is conserved, $\dot{\bf K}=0$. The CM motion is coupled to the internal motion via non-trivial electric fields ${\bf E}_{R}(\dot{\ta_1},\,\dot{\ta_2})$, ${\bf E}_{1}({\bf R},\,{\ta}_2,\,\dot{ \ta}_2)$ and ${\bf E}_{2}({\bf R},\,{\ta}_1,\,\dot{ \ta}_1)$ which vanish at $e_{c1}=e_{c2}=0$. Also, the variables $\tau_1$ and $\tau_2$ themselves are strongly coupled with each other via the Coulomb potential. We emphasize that even in the Born-Oppenheimer approximation, where $M \rightarrow \infty$, the CM and relative variables continue to be coupled.

In the case of a neutral system ($Q=0$) at rest (${\bf K}=0$) with two identical particles ($e_{c1}=0$)

\[
e_1 \ = \ e_2 \ \equiv e \ , \qquad m_1 \ = \ m_2 \ \equiv \ m \ ,
\]

the electric fields ${\bf E}_{1}$, ${\bf E}_{2}$ disappear. The corresponding quantum neutral system was shown to possess exact factorizable solutions in the absence of the Coulomb interaction \cite{Simonov}. In the present work, the origin of such solvability becomes more transparent.

To the best of the knowledge of the present authors, the equations (\ref{Eq}) and the associated classical and quantum Hamiltonian (\ref{HC}) have not been discussed in the literature in full generality.

\vspace{0.15cm}

The above analysis of the equations of motion indicates the existence of three important particular cases, namely

(i)\ $Q=0$ (neutral system): the components of ${\bf K}$ are in involution, $\{ K_x, K_y \}=0$\ , and a pseudo-separation of the CM motion can be achieved \cite{Avron}.
%

%
(ii)\ $e_{c1}=e_{c2}=0$ (particles with the same charge-to-mass ratio): In the equations (\ref{Eq}) the center of mass variable can be separated out.

(iii)\ $Q=0,\, e_2=e_3;\,m_2=m_3$ (neutral system at rest ${\bf K}=0$ with two identical particles): In the equations (\ref{Eq}), a certain number of linear terms (electric fields) can be removed.

\vspace{0.2cm}


\section{SPECIAL TRAJECTORIES}
\label{ST}

In this section special trajectories where the distances between the particles remain unchanged are presented. We focus on the classification of the initial data associated with these periodic orbits and the corresponding integrals and \emph{particular} constants of motion. A precise distinction between integrals and \emph{particular} constants is in order.

\subsection*{Integrals and \emph{particular} constants of motion}

A function ${\cal I}={\cal I}({\bo},\,{\bf p})$, defined in the phase space, such that

\[
\{ H,\,{\cal I} \} \ = \  F({\bo},\,{\bf p}) \ \neq \ 0 \ ,
\]

is called a \emph{particular} constant of motion\cite{Turbiner:2013p} if there exists, within the domain where the problem is defined, a trajectory $\Sigma$ for which $F\,|_\Sigma=0$. Evaluated along $\Sigma$, the Poisson bracket vanishes $\{ H,\,{\cal I} \} =0$ and, consequently, ${\cal I}\,|_\Sigma$ is conserved. Of course, this implies the choice of specific initial conditions.

In the case of an integral of motion, like the Pseudomomentum $\bf K$ (\ref{pseudomomentum}), the Poisson bracket $\{ H,\,{\bf K} \} =0$ is identically zero and thus independent of the initial conditions.

For the three body problem we will show that along the special trajectories: (1) unlike the generic case (\ref{AlgebraInt}), the integrals $K_x,\,K_y,\,L^{total}_z $ are Poisson commuting invariants, and (2) \emph{particular} constants of motion occur.

\subsection{CONFIGURATION I}

\begin{center}
\begin{figure}[h]
\hspace{1.5cm} \includegraphics[scale=0.3]{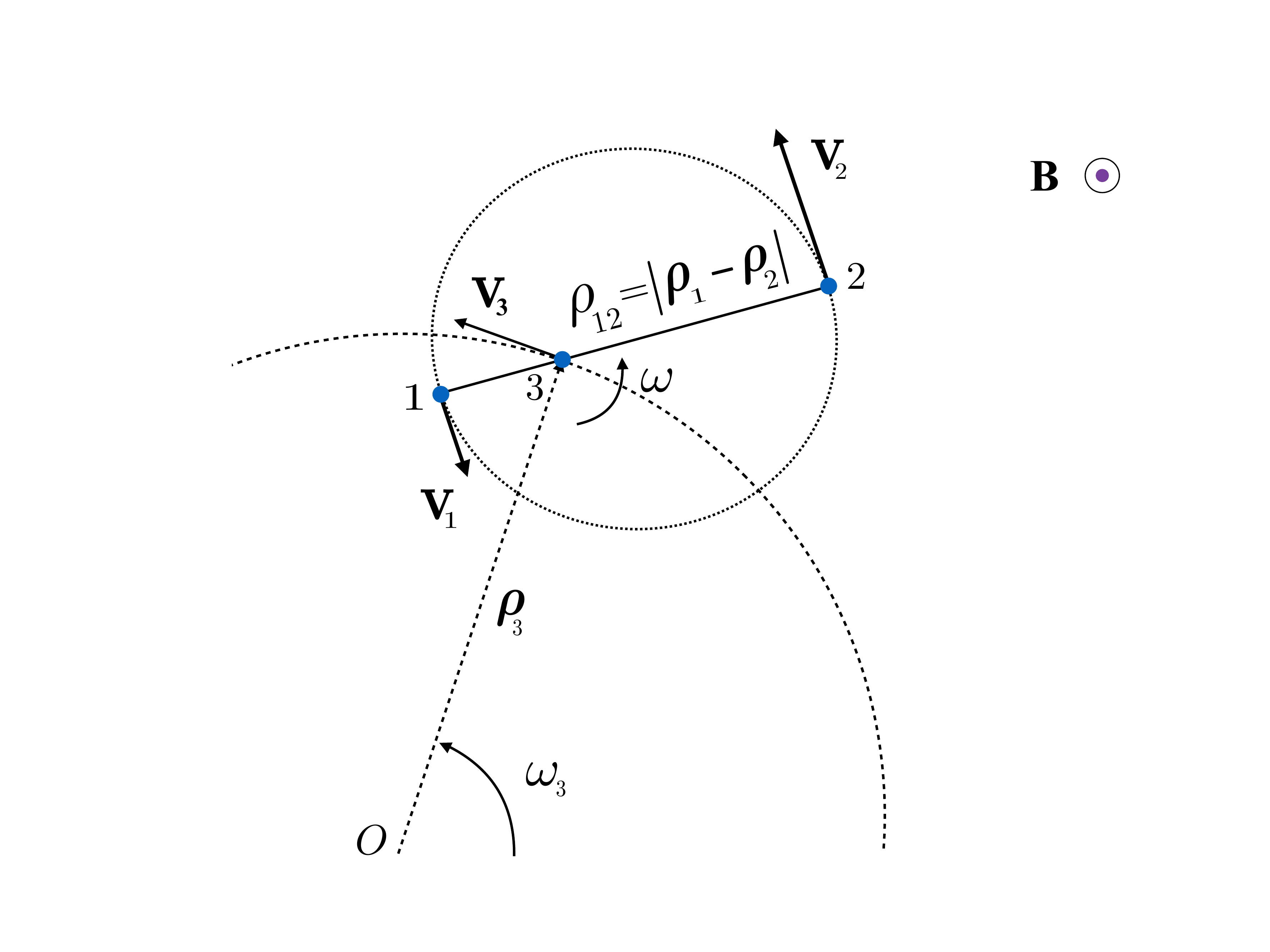}
\caption{ \label{Config1}Configuration I. Two charges $e_1$ and $e_2$ rotate, with angular frequency $\omega$ and opposite velocities, around $e_3$ which by itself rotates with frequency $\omega_3$. At $\rm{v}_3=0$ (the inner charge at rest), these special trajectories are realized by the three electron, neutral and Helium-like systems. For $\rm{v}_3 \neq 0$, this configuration appears only for charges of the same sign.  }
\end{figure}
\end{center}

%
The Configuration I corresponds to the case when two charges $e_1$ and $e_2$ rotate, with the same angular frequency $\omega$ and opposite velocities, around the third charge $e_3$. In its own the charge $e_3$ rotates with frequency $\omega_3$ around a fixed point.
This Configuration I is the superposition of two independent circular motions, it is presented in Fig. \ref{Config1}. The explicit form, as a function of time, of the corresponding special trajectories is given by

\begin{equation}
\begin{aligned}
& {\bo}_1(t) \ = \ \frac{{\rm v}_1}{\omega}\,(\,\cos \omega t  ,\, - \sin \omega t ) \ + \ {\bo}_3(t) \ ,
\\ & {\bo}_2(t) \ = \ -\frac{{\rm v}_2}{\omega}\,(\,\cos \omega t ,\, - \sin \omega t) \ + \ {\bo}_3(t) \ ,
\\ & {\bo}_3(t) \ = \ \frac{{\rm v}_3}{\omega_3}\,(\,\cos \omega_3 t ,\, - \sin \omega_3 t ) \ ,
\label{tC1}
\end{aligned}
\end{equation}

where ${\rm v}_1>0, {\rm v}_2>0, {\rm v}_3\geq0$, $\omega$ and $\omega_3$ are real parameters to be determined by the Newton equations (\ref{NEsep}).
For these trajectories (\ref{tC1}) the relative distances between the particles remain unchanged during time evolution
\begin{equation}
\label{rhoIa}
|\bo_1-\bo_2| \ =\  \frac{{\rm v}_1+{\rm v}_2}{\omega}\ , \qquad \ |\bo_1-\bo_3| \ =\  \frac{{\rm v}_1}{\omega} \ , \qquad \ |\bo_2-\bo_3| \ =\  \frac{{\rm v}_2}{\omega} \ ,
\end{equation}
hereafter without loss of generality we assume $\omega>0$.
Substituting (\ref{tC1}) in (\ref{NEsep}) we obtain the following set of algebraic equations
\begin{equation}
{\rm v}_3\,(B\,e_1-m_1\,\omega_3)\,=\,0 \ , \quad {\rm v}_3\,(B\,e_2-m_2\,\omega_3) \,=\,0 \ , \quad  {\rm v}_3\,(B\,e_3-m_3\,\omega_3) \,=\,0 \ ,
\label{t1}
\end{equation}
\begin{equation}
e_3\bigg(\frac{e_1}{{\rm v}_1^2}-\frac{e_2}{{\rm v}_2^2}\bigg)\omega^2=0 \ ,
\label{t2}
\end{equation}
\begin{equation}
B\,e_1\,{\rm v}_1-\omega\bigg[m_1{\rm v}_1+\frac{e_1(e_2{\rm v}_1^2+e_3({\rm v}_1+{\rm v}_2)^2)\,\omega}{{\rm v}_1^2\,({\rm v}_1+{\rm v}_2)^2}\bigg]\ =\ 0 \ ,
\label{t3}
\end{equation}
\begin{equation}
B\,e_2\,{\rm v}_2-\omega\bigg[m_2{\rm v}_2+\frac{e_2(e_1{\rm v}_2^2+e_3({\rm v}_1+{\rm v}_2)^2)\,\omega}{{\rm v}_2^2\,({\rm v}_1+{\rm v}_2)^2}\bigg]\ =\  0 \ .
\label{t4}
\end{equation}

We consider fixed charges and masses, the goal is to find ${\rm v}_i$ ($i=1,2,3$), the magnetic field $B$ and the two angular frequencies $\omega$ and $\omega_3$ such that the equations (\ref{t1})-(\ref{t4}) are satisfied. The solutions of Eqs. (\ref{t1})-(\ref{t4}) determine completely the initial conditions for which the Configuration I occurs.

Now, notice that in (\ref{t1}) all the equations are proportional to $\rm v_3$ while Eqs. (\ref{t2})-(\ref{t4}) do not depend on $\rm v_3$ at all. Thus, we distinguish two cases ${\rm v}_3=0$ and ${\rm v}_3\neq0$.

\subsubsection{Case ${\rm v}_3=0$}

This corresponds to the situation when the third charge ($e_3$) is at rest, ${\rm v}_3=0$ and thus $\bo_3(t)=0$, while the two charges ($e_1,e_2$) move around it in clockwise (or counterclockwise) direction with frequency $\omega$ and antiparallel velocities. For ${\rm v}_3=0$, the equations (\ref{t1}) are satisfied trivially. The Eq. (\ref{t2}) implies $e_1\,e_2>0$ (charges of the same sign) and it leads to the following expression for ${\rm v}_2$

\begin{equation}\label{velCI}
{\rm v}_2 \ = \ {\rm v}_1\,\sqrt{\frac{e_2}{e_1}} \ .
\end{equation}

For arbitrary magnetic field $B$, to determine a closed analytical expression for ${\rm v}_1$ solution of the remaining equations (\ref{t3})-(\ref{t4}) is not only a difficult task but unnecessary. In fact, we can indicate the value of magnetic field $B\equiv B_I$ for which these special trajectories occur, it is derived from the compatibility condition of the Eqs. (\ref{t3})-(\ref{t4})

\begin{equation}
\label{BI}
B_I \ = \   \frac{(e_2\,m_1-e_1\,m_2)\,r\,{(1+r)}^2\,(m_1-m_2\,r)\,{\rm v}_1^3}{e_1\,{(e_1-r\,e_2)}^2\,(e_2+e_3\,{(1+r)}^2)   }\ \neq \ 0 \ ,
\end{equation}

where we used (\ref{velCI}) and $r=\sqrt{\frac{e_2}{e_1}}$. For the magnetic field $B_I$, the frequency $\omega$ is

\[
\omega \ = \  \frac{(e_2\,m_1-e_1\,m_2)\,r\,{(1+r)}^2\,{\rm v}_1^3}{e_1\,{(e_1-r\,e_2)}\,(e_2+e_3\,{(1+r)}^2)   }  \ > \ 0 \ .
\]

Notice that both $B_I$ and $\omega$ vanish at $(e_2\,m_1-e_1\,m_2) = 0$, i.e. at $e_{c1}=0$. Therefore the case $e_{c1}=0$ should be considered separately.

\subsubsection*{Two particles with equal Larmor frequency ($e_{c1}=0$)}

Direct analysis of (\ref{t2})-(\ref{t4}) shows that for $e_{c1} = 0$ there exist special trajectories for two identical particles only. More precisely, for two identical particles $e_1=e_2=e$, $m_1=m_2=m$, and $\rm{v}_1=\rm{v}_2=\rm{v}$, the Eqs. (\ref{t3})-(\ref{t4}) do coincide. Then, the Eqs. (\ref{t3})-(\ref{t4}) reduce to a single equation
\begin{equation}
4\,B\,e\,{\rm v}^3-\omega\,[4\,m\, {\rm v}^3+e\,(e+4\, e_3)\,\omega]\ =\ 0 \ .
\label{rel1}
\end{equation}
From Eqs. (\ref{rel1}) and (\ref{rhoIa}) we immediately find
\begin{equation}
  {\rm v}=\frac{e\,B\,\rho_{12}}{4\,m}\bigg(1\pm \sqrt{1-\frac{8\,m\,(e+4\,e_3)}{e\,B^2\,\rho_{12}^3}}\bigg)\ .
\label{ec2}
\end{equation}

Thus, for given $\rho_{12}$ there exist two different initial velocities $ \rm{v}$ leading to the same circular trajectory presented in Fig. \ref{Config1}. However, for given ${\rm v}$ there exists a single circular trajectory with a certain $\rho_{12}$. It corresponds to rotation with frequency equal to $\frac{{2\,\rm v}}{\rho_{12}}$. It is interesting that for a given magnetic field there exists a minimal circular trajectory with $\rho_{12}=\rho_{min}={ (  \frac{8\,m}{e\,B^2}(e+4\,e_3))}^{\frac{1}{3}}$ (when the square root in (\ref{ec2}) vanishes).

\subsubsection*{Conserved quantities ($\rm v_3=0$)}

At $\rm v_3=0$, the system is \emph{particularly superintegrable}. Evaluated along the special trajectories (\ref{tC1}), the six quantities $({\cal H},\,{\bf K}^2, {\bf L}^{total}_z,\,\ell_{z_3},\,T_{1},\,T_{2})$ are in involution where $T_i = \frac{{({\bf p}_i-e_i\,{\bf A}_{{\bo}_{i}})}^2}{2\,m_i}$, $i=1,2$\ . Moreover, the function ${\cal I}=\ta_{1}\cdot {\bf p}_{\ta_1}$ is an extra \emph{particular} constant of motion. We emphasize that the three quantities $({\cal H},\,{\bf K}^2, {\bf L}^{total}_z)$ are integrals, they are conserved for any trajectory (arbitrary initial conditions) while $(\ell_{z_3},\,T_{1},\,T_{2},\,{\cal I})$ are constants only for special ones (constraint on initial conditions (\ref{t1})-(\ref{t4})\,).

Explicitly, they take the values

{\footnotesize
\begin{equation}
\begin{aligned}
& {\cal H} \  =\  \frac{{\rm v}_1^2}{2}  \bigg[  m_1+ m_2\,r^2  + \frac{2\,e_1\,r^2\,(1+r)|m_2-m_1\,r^2|}{|e_1\,(r^3-1)\,(e_1\,r^2+e_3(1+r))|}(e_1\,r+e_3(1+r))
 + \frac{2\,e_1\,e_3\,{(1+r)}^2|m_2\,r-m_1\,r^3|}{|e_1\,(r^3-1)\,(e_1\,r^2+e_3(1+r))|}       \bigg] \ ,
\\ &
{\bf K}^2   \ = \  0 \ ,
\\ &
{L}^{total}_z   = \frac{e_1(r^3-1)(e_1r^2+e_3{(1+r)}^2)[B_I e_1^2(r^3-1-r^4+r^7)(e_1r^2+e_3{(1+r)}^2)
   + 2r{(1+r)}^2{\rm v}_1^3(m_1r^2-m_2)(m_2r^2+m_1)  ]}{2r^2{(1+r)}^4{\rm v}_1^4{(m_2-m_1r^2)}^2} \ ,
\\ &
 T_{1}\ =\ \frac{1}{2}m_1\,{\rm v}_1^2  \ ,\qquad T_{2}\ =\ \frac{1}{2}m_1\,r^2\,{\rm v}_1^2 \  , \qquad \ell_{z_3}   \ =\ 0 \  , \qquad {\cal I} \ =\ 0 \ ,
\end{aligned}
\end{equation}
}

where ${\rm v}_1>0,\,e_{c1}\neq 0$, $B_I = \frac{r\,{(1+r)}^2\,(m_1-m_2\,r)(m_1\,r^2-m_2)\,{\rm v}_1^3}{e_1^2\,{(r^3-1)}^2\,(e_1\,r^2+e_3\,{(1+r)}^2)}$ and $r^2 = \frac{e_2}{e_1}$\ .

\subsubsection{Case ${\rm v}_3\neq0$}

\begin{center}
\begin{figure}[h]
\includegraphics[scale=0.5]{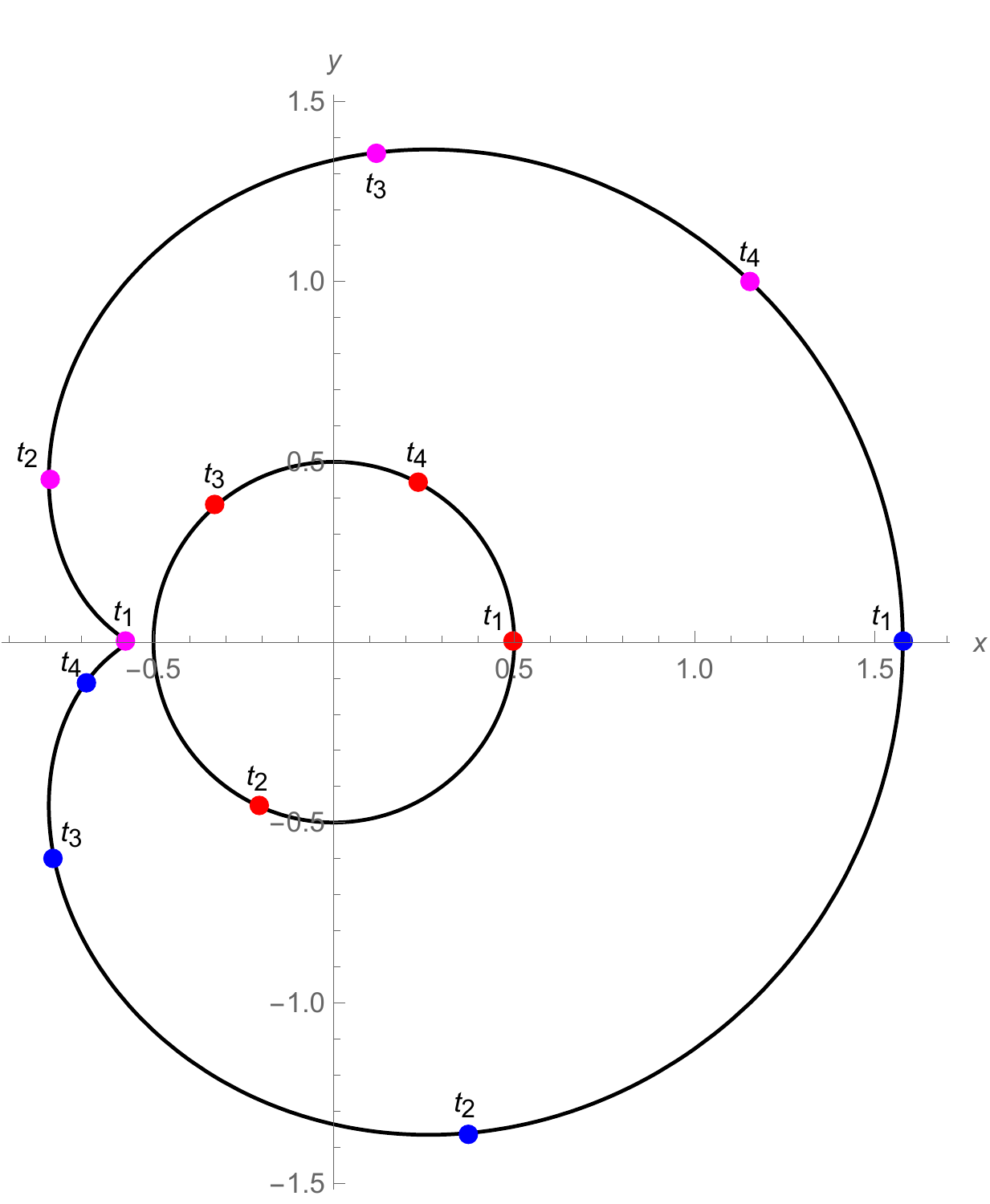}
\caption{\label{3est}Configuration I ($\rm v_3\neq0$): special trajectories for a three electron system, $e_1=e_2=e_3=-1$ and $m_1=m_2=m_3=1$. The numerical solutions of the Newton equations (\ref{NEsep}) are displayed for the values $B=-2$, $\rm v_1=\rm v_2=\sqrt[3]{\frac{5}{4}}$, $\rm v_3=1$ and $\omega=1,\,\omega_3=2$ obtained by solving the algebraic equations (\ref{w3})-(\ref{eq1e2}). The successive positions of the third charge, which moves in a circular trajectory, are indicated by red points at different times $t_1<t_2<t_3<t_4$. Similarly, the locations of the first charge (blue points) and the second one (magenta points) are presented. The relative distances between particles remain constant during time evolution.   }
\end{figure}
\end{center}

For ${\rm v}_3\neq0$ the Eq. (\ref{t1}) implies that both coupling charges $e_{c1}=e_{c2}=0$ vanish, namely, the three charges must possess the same charge-to-mass ratio

\begin{equation}
\frac{e_1}{m_1}\ =\ \frac{e_2}{m_2}\ =\  \frac{e_3}{m_3}\ \equiv \ \alpha \ .
\label{cargamasa}
\end{equation}
Substituting (\ref{tC1}) into the Newton equations of motion (\ref{NEsep}) we find the algebraic equations

\begin{equation}
\omega_3 \ = \ B\,\alpha \ ,
\label{w3}
\end{equation}
\begin{equation}
\frac{e_1}{{\rm v}_1^2}=\frac{e_2}{{\rm v}_2^2} \ ,
\label{equalcharges}
\end{equation}
\begin{equation}
B\,\alpha \ - \  \omega \bigg(1 + \frac{e_1(e_2{\rm v}_1^2+e_3({\rm v}_1+{\rm v}_2)^2)\,\omega}{m_1\,{\rm v}_1^3({\rm v}_1+{\rm v}_2)^2}\bigg) \ =\ 0\ ,
\label{eq1e1}
\end{equation}
\begin{equation}
B\,\alpha \ -  \  \omega \bigg(1 + \frac{e_2(e_1{\rm v}_2^2+e_3({\rm v}_1+{\rm v}_2)^2)\,\omega}{m_2\,{\rm v}_2^3({\rm v}_1+{\rm v}_2)^2}\bigg) \ =\ 0 \ .
\label{eq1e2}
\end{equation}
In particular, the condition (\ref{w3}) implies that the motion of the charge $e_3$ corresponds to that of a free particle in a constant magnetic field $B$, thus it rotates on a circular path with frequency $\omega_3=B\,\alpha$ and arbitrary $\rm{v}_3>0$. Moreover, the Eqs. (\ref{eq1e1})-(\ref{eq1e2}) are satisfied for two identical particles $e_1=e_2=e$, $m_1=m_2=m$, and $\rm{v}_1=\rm{v}_2=\rm{v}$ only. The corresponding motion was already described in detail in the previous section, see (\ref{ec2}). By virtue of (\ref{cargamasa}), all three charges must be of the same sign and $(e_3\,m-e\,m_3)=0$. An example is presented in Fig. \ref{3est}.

\subsubsection*{Conserved quantities $\rm v_3\neq0$}

For the case $\rm v_3\neq0$, two particles are identical $e_1=e_2=e$, $m_1=m_2=m$ and $\rm{v}_1=\rm{v}_2=\rm{v}$. The system is again \emph{particularly superintegrable}. Along any special trajectory the six quantities $({\cal H},\,{\bf K}^2,\,{\bf L}^{total}_z,\,\ell_{z_3},\,T_{3},\,k_{3})$ where $k_3=({\bf p}_3+e_3\,{\bf A}_{{\bo}_{3}})\cdot {\hat {\bf x}}$ are in involution. Again, the function ${\cal I}=\ta_{1}\cdot {\bf p}_{\ta_1}$ is an extra \emph{particular} constant of motion.

Explicitly, they take the values

\begin{equation}
\begin{aligned}
& {\cal H} \  =\  \frac{m_3\,\rm v_3^2}{2}\ + \ \frac{e^2\,B^2\,\rho_{12}^2}{16\,m}{\bigg[ 1 \pm  \sqrt{1- \frac{8\,m\,(e+4\,e_3)}{e\,B^2\,\rho_{12}^3}}  \bigg]}^2 \
 + \ \frac{e\,(e+4\,e_3)}{\rho_{12}}
 \ , \\ &
{\bf K}^2 \ = \ 0
 \ , \\ &
{L}^{total}_z \ =  \ \frac{e\,B\,\rho_{12}^2}{4} - \frac{(2\,m+m_3)\,m_3\,\rm v_3^2}{2\,e_3\,B} - \frac{e\,B\,\rho_{12}^2}{4}\bigg(1\pm \sqrt{1-\frac{8\,m\,(e+4\,e_3)}{e\,B^2\,\rho_{12}^3}}\bigg)
 \ , \\ &
 \ell_{z_3} \ = \ -\frac{m_3^2\,\rm v_3^2}{2\,e_3\,B}
 \ , \\ &
T_{3} \ = \ \frac{m_3\,\rm v_3^2}{2}
 \ , \\ &
k_{3} \ = \ 0 \ , \qquad \ {\cal I} \ = \ 0
 \ ,
\end{aligned}
\end{equation}

where $\rho_{12}$ is taken from (\ref{ec2}). The allowed values of the magnetic field are given by $B^2\geq \frac{8\,m\,(e+4\,e_3)}{e\,\rho_{12}^3}$.

\hskip 1cm
%


\begin{center}
\subsection{CONFIGURATION II}
\end{center}

\bigskip

\begin{center}
\begin{figure}[h]
\hspace{1.5cm} \includegraphics[scale=0.2]{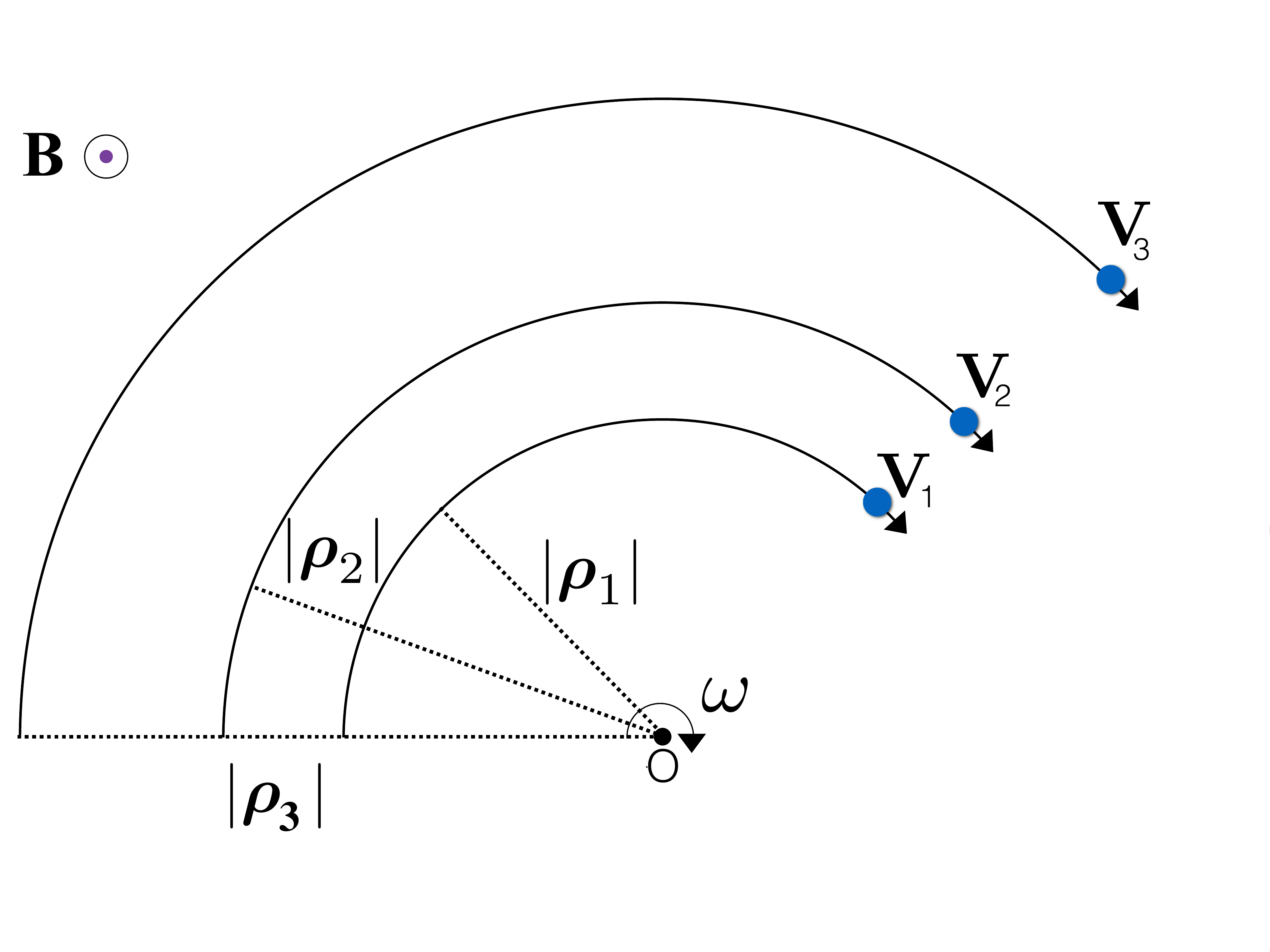}
\caption{\label{Config2}Configuration II. Three particles rotate clockwise (or counterclockwise) in phase, around a fixed common center, with the same frequency $\omega$. This Configuration is not admitted for three-electron systems. }
\end{figure}
\end{center}

The Configuration II corresponds to three particles rotating clockwise (or counterclockwise) in phase, around a fixed common center, all with the same frequency $\omega$. This Configuration II is presented in Fig. \ref{Config2}.

As a function of time, the trajectories are given by

\begin{equation}
\begin{aligned}
& {\bo}_1(t) \ = \ \frac{{\rm v}_1}{\omega}\,(\,\cos \omega t  ,\, - \sin \omega t )  \ ,
\\ & {\bo}_2(t) \ = \ \frac{{\rm v}_2}{\omega}\,(\,\cos \omega t ,\, - \sin \omega t)  \ ,
\\ & {\bo}_3(t) \ = \ \frac{{\rm v}_3}{\omega}\,(\,\cos \omega t ,\, - \sin \omega t ) \ ,
\label{tC2}
\end{aligned}
\end{equation}

where without losing generality we have assumed
\[
\rm{v}_3>\rm{v}_2>\rm{v}_1>0 \ ; \qquad \omega >0 \ .
\]
The concrete value of these quantities, as in the previous case, is determined by the Eqs. (\ref{NEsep}). The constant relative distances between the particles read

\begin{equation}
|{\bo}_1-{\bo}_2|=\frac{\rm{v}_2-\rm{v}_1}{\omega}\, , \,\quad
|{\bo}_1-{\bo}_3|=\frac{\rm{v}_3-\rm{v}_1}{\omega}\, , \,\quad
|{\bo}_2-{\bo}_3|=\frac{\rm{v}_3-\rm{v}_2}{\omega}\, , \,\quad
\end{equation}
they remain unchanged during time evolution.

Putting (\ref{tC2}) into the Newton equations (\ref{NEsep}) we arrive to three algebraic equations
\begin{equation}
\label{eqC21}
B\,e_1{\rm v}_1-m_1\,{\rm v}_1\,\omega + e_1\,\omega^2\bigg(\frac{e_2}{({\rm v}_1-{\rm v}_2)^2}+\frac{e_3}{({\rm v}_1-{\rm v}_3)^2}\bigg)\ = \ 0 \ ,
\end{equation}
\begin{equation}
\label{eqC22}
B\,e_2{\rm v}_2-m_2\,{\rm v}_2\,\omega + e_2\,\omega^2\bigg(\frac{e_3}{({\rm v}_2-{\rm v}_3)^2}-\frac{e_1}{({\rm v}_1-{\rm v}_2)^2}\bigg)\ =\ 0 \ ,
\end{equation}
\begin{equation}
\label{eqC23}
B\,e_3{\rm v}_3-m_3\,{\rm v}_3\,\omega - e_3\,\omega^2\bigg(\frac{e_1}{({\rm v}_1-{\rm v}_3)^2}+\frac{e_2}{({\rm v}_2-{\rm
v}_3)^2}\bigg)\ =\ 0 \ .
\end{equation}

From (\ref{eqC21})-(\ref{eqC23}) it follows that for particles with equal charge to mass ratio
\[
\frac{e_1}{m_1} \ = \ \frac{e_2}{m_2}\ = \ \frac{e_3}{m_3} \ = \ \alpha \ ,
\]
($\alpha \neq 0$ is a real constant) the Configuration II does not occur. Therefore, three electrons cannot perform this Configuration II.

\vspace{0.2cm}

Now, the sum of the equations (\ref{eqC21})-(\ref{eqC23}) gives the following relation between the magnetic field $B$ and the frequency $\omega$

\begin{equation}
\label{relC2}
\omega \ = \  \bigg(\frac{e_1 \,{\rm v}_1+e_2 \,{\rm v}_2+e_3\, {\rm v}_3  }{m_1 \,{\rm v}_1+m_2 \,{\rm v}_2+m_3 \,{\rm v}_3 }\bigg)\,B \ .
\end{equation}

The above relation (\ref{relC2}) is a necessary (not sufficient) condition for the existence of special trajectories in the Configuration II.  Substituting (\ref{relC2}) into the equation (\ref{eqC22}) and solving for $B \equiv B_{II}$ we obtain

\begin{equation}
\label{BC2}
B_{II} \ = \  \frac{\left({\rm v}_1-{\rm v}_2\right){}^2 {\rm v}_2 \left({\rm v}_2-{\rm v}_3\right){}^2 \left(m_1 {\rm v}_1+m_2 {\rm v}_2+m_3 {\rm v}_3\right) \left(e_2 \left(m_1 {\rm v}_1+m_3 {\rm v}_3\right)-m_2 \left(e_1 {\rm v}_1+e_3 {\rm v}_3\right)\right)}{e_2 \left(e_1 \left({\rm v}_2-{\rm v}_3\right){}^2-e_3 \left({\rm v}_1-{\rm v}_2\right){}^2\right) \left(e_1 {\rm v}_1+e_2 {\rm v}_2+e_3 {\rm v}_3\right){}^2} \ .
\end{equation}

From a physical point of view we have to impose $B_{II}\neq0$ and finite.

In general, substituting (\ref{BC2}) into (\ref{eqC23}) we get a multivariate polynomial equation of sixth order in the variables ${\rm v}_1,\,{\rm v}_2,\,{\rm v}_3$

\begin{equation}
\label{Pol2}
\mathbb{P}_6({\rm v}_1,\,{\rm v}_2,\,{\rm v}_3) \ \equiv \ \sum_{i+j+k=6} a_{ijk}\,{\rm v}_1^i\,{\rm v}_2^j\,{\rm v}_3^k \ = \ 0 \ ,
\end{equation}

where the coefficients $a_{ijk}=a_{ijk}(e_1,e_2,e_3,m_1,m_2,m_3)$ are completely determined by the system we deal with, (see Appendix). For fixed charges and masses in (\ref{Pol2}), it is not possible to write its general solution analytically, meaning a relation of the form ${\rm v}_1=(e_1,e_2,e_3,m_1,m_2,m_3;{\rm v}_2,{\rm v}_3)$ such that (\ref{Pol2}) is satisfied. However, we can do so in the following physically relevant systems.


\subsubsection*{Particles with special charge to mass ratio}

When the charge to mass ratio of the particle $e_1$ is equal to the ratio of a composite particle (formed by the other charges), namely
\begin{equation}
\label{Wf}
 \frac{e_1}{m_1} \ - \  \frac{(e_2 \,{\rm v}_2+e_3 \, {\rm v}_3)}{\left(m_2 \,{\rm v}_2+m_3\, {\rm v}_3\right)}  \ = \ 0 \ ,
\end{equation}

( $\frac{e_1}{m_1}\neq \frac{e_3}{m_3}$) then the polynomial (\ref{Pol2}) simplifies. In this case we obtain effectively a quadratic polynomial in ${\rm v}_1$ solutions of which are given by
\begin{equation}
\label{v1Wf}
{\rm v}_1 \ = \ \frac{e_1 {\rm v}_3 \left(m_2 {\rm v}_2+m_3 {\rm v}_3\right)+e_3 m_1 \left({\rm v}_2^2-{\rm v}_3^2\right) \pm \left({\rm v}_2-{\rm v}_3\right) \sqrt{e_3 m_1 {\rm v}_2 \left(e_3 m_1 {\rm v}_3-e_1 \left(m_2 {\rm v}_2+m_3 {\rm v}_3\right)\right)}}{e_3 m_1 \left({\rm v}_2-{\rm v}_3\right)+e_1 \left(m_2 {\rm v}_2+m_3 {\rm v}_3\right)} \ ,
\end{equation}

and they give rise to the Configuration II. There exist special values of velocities and masses for which the neutral system $Q=0$ occurs in this Configuration II. The corresponding magnetic field $B$ is given by (\ref{BC2}) with ${\rm v}_1$ taken from (\ref{v1Wf}).


\subsubsection*{Helium-like system}

Here we consider in detail the physically important case of a neutral system

\[
Q \ = \ e_1 \ + \ e_2 \ + \ e_3  \ = \ 0 \ ,
\]
with two identical particles (Helium-like system)
\[
e_2 \ = \ e_3 \ \equiv \ e \ , \qquad m_2 \ = \ m_3 \ \equiv \ m \ , \qquad \frac{e_1}{m_1}\neq \frac{e}{m} \ .
\]

For any initial ${\rm v}_2 $ and ${\rm v}_3$ such that

\begin{equation}
\label{2PF}
{\rm v}_3 \ \geq \ \lambda \ > \ {\rm v}_2  \ > \  0 \ ,
\end{equation}

where $\lambda$ obeys the cubic equation

\begin{equation}
\begin{aligned}
&  \lambda^3  \ - \ 117 \,{\rm v}_{2} \,\lambda^2 \  -  \  81\, {\rm v}_{2}^2\, \lambda \  - \  27\, {\rm v}_{2}^3 \ = \ 0 \ ,
\end{aligned}
\end{equation}

thus $\lambda=\lambda({\rm v}_2)$, there exists an initial velocity ${\rm v}_1$ for which the Configuration II occurs. This ${\rm v}_1={\rm v}_1({\rm v}_2,\,{\rm v}_3)$ obeys the following quartic equation

\begin{equation}
\begin{aligned}
&  {\rm v}_{1}^4\,({\rm v}_3+{\rm v}_2)\ -  \  2\,{\rm v}_{1}^3\,( {\rm v}_3^2 + 2 {\rm v}_3 {\rm v}_2+  {\rm v}_2^2)
\ + \  {\rm v}_{1}^2\,(3 {\rm v}_3^3 - {\rm v}_3^2 {\rm v}_2 + 11 {\rm v}_3 {\rm v}_2^2 - {\rm v}_2^3)
\\ & +  2\,{\rm v}_{1}\,(  3 {\rm v}_3^3 {\rm v}_2 - 2 {\rm v}_3^2 {\rm v}_2^2 - 5 {\rm v}_3 {\rm v}_2^3 + 2 {\rm v}_2^4-2 {\rm v}_3^4)
+   (2 {\rm v}_3^5 - 4 {\rm v}_3^4 {\rm v}_2 + 3 {\rm v}_3^3 {\rm v}_2^2 - {\rm v}_3^2 {\rm v}_2^3 + 4 {\rm v}_3 {\rm v}_2^4 - 2 {\rm v}_2^5 )\ = \ 0 \ .
\end{aligned}
\end{equation}

The corresponding magnetic field $B$ is given by
\[
B_{II}^{Helium-system} \ = \ \frac{\left(2 m+m_1\right) {\rm v}_1 {\rm v}_2 \left({\rm v}_1-{\rm v}_2\right){}^2  \left({\rm v}_2-{\rm v}_3\right){}^2 \left(m_1 {\rm v}_1+m \left({\rm v}_2+{\rm v}_3\right)\right)}{e^3 \left(2 {\rm v}_1-{\rm v}_2-{\rm v}_3\right){}^2 \left({\rm v}_1^2-2 {\rm v}_2 {\rm v}_1+3 {\rm v}_2^2+2 {\rm v}_3^2-4 {\rm v}_2 {\rm v}_3\right)} \ ,
\]
and the frequency takes the form
\[
\omega \ = \  \frac{\left(2 m+m_1\right) {\rm v}_1 \left({\rm v}_1-{\rm v}_2\right){}^2 {\rm v}_2 \left({\rm v}_2-{\rm v}_3\right){}^2}{e^2 \left(2 {\rm v}_1-{\rm v}_2-{\rm v}_3\right) \left({\rm v}_1^2-2 {\rm v}_2 {\rm v}_1+3 {\rm v}_2^2+2 {\rm v}_3^2-4 {\rm v}_2 {\rm v}_3\right)} \ .
\]

Therefore, a two-parametric family of special trajectories occur. They are functions of the parameters ${\rm v}_2$ and ${\rm v}_3$, see (\ref{2PF}).

\subsubsection*{Conserved quantities}

For the Configuration II shown in Fig. (\ref{Config2}), the system is again \emph{particularly superintegrable}. Along any special trajectory (\ref{tC2}) the six quantities $({\cal H},\,{\bf K}^2,\,{\bf L}^{total}_z,\,\ell_{z_2},\,T_{1},\,T_{2})$ are in involution. The function ${\cal I}=\ta_{1}\cdot {\bf p}_{\ta_1}$ is an extra \emph{particular} constant of motion as well.

Explicitly, the integrals and \emph{particular} constants take the values
{\small
\begin{equation}
\begin{aligned}
\label{IntMCII}
& {\cal H} \  =\ \frac{1}{2} \left[\frac{2\, B_{II}\, \left(\frac{e_2 e_3}{{\rm v}_2-{\rm v}_3}+ \frac{e_1\,e_2}{{\rm v}_1-{\rm v}_2}+\frac{e_1\,e_3}{{\rm v}_1-{\rm v}_3}\right) \left(e_1\, {\rm v}_1+e_2\, {\rm v}_2+e_3\, {\rm v}_3\right)}{m_1\, {\rm v}_1+m_2 \,{\rm v}_2+m_3\, {\rm v}_3}+m_1 \,{\rm v}_1^2+m_2\, {\rm v}_2^2+m_3 \,{\rm v}_3^2\right]
 \ , \\ &
{\bf K}^2 \  = \ 0 \ , \\ &
{L}^{total}_z \  = \ \frac{\left(m_1 {\rm v}_1+m_2 {\rm v}_2+m_3 {\rm v}_3\right){}^2 }{2\, B_{II}\, \left(e_1 {\rm v}_1+e_2 {\rm v}_2+e_3 {\rm v}_3\right){}^2}\left[e_1 {\rm v}_1^2+e_2 {\rm v}_2^2+e_3 {\rm v}_3^2-\frac{2 \left(e_1 {\rm v}_1+e_2 {\rm v}_2+e_3 {\rm v}_3\right) \left(m_1 {\rm v}_1^2+m_2 {\rm v}_2^2+m_3 {\rm v}_3^2\right)}{m_1 {\rm v}_1+m_2 {\rm v}_2+m_3 {\rm v}_3}\right]\,, \\ &
\ell_{z_2} \  =  \  \frac{{\rm v}_2^2 \left(m_1 {\rm v}_1+m_2 {\rm v}_2+m_3 {\rm v}_3\right) \left[\ e_2 \left(m_1 {\rm v}_1-m_2 {\rm v}_2+m_3 {\rm v}_3\right)-2 m_2 \left(e_1 {\rm v}_1+e_3 {\rm v}_3\right)\ \right]}{2 \,B_{II}\, \left(e_1 {\rm v}_1+e_2 {\rm v}_2+e_3 {\rm v}_3\right){}^2}   \   , \\ &
T_i \  =  \  \frac{m_i\,{\rm v}_i^2}{2} \ , \qquad (i=1,2) \ , \\ &
{\cal I} \  =  \  0 \ ,
\end{aligned}
\end{equation}
}

where $B_{II}$ is given in (\ref{BC2}) and ${\rm v}_1,\,{\rm v}_2,\,{\rm v}_3$ are solutions of the polynomial equation (\ref{Pol2}).


\subsection{CONFIGURATION III}

\begin{center}
\begin{figure}[h]
\hspace{1.0cm} \includegraphics[scale=0.2]{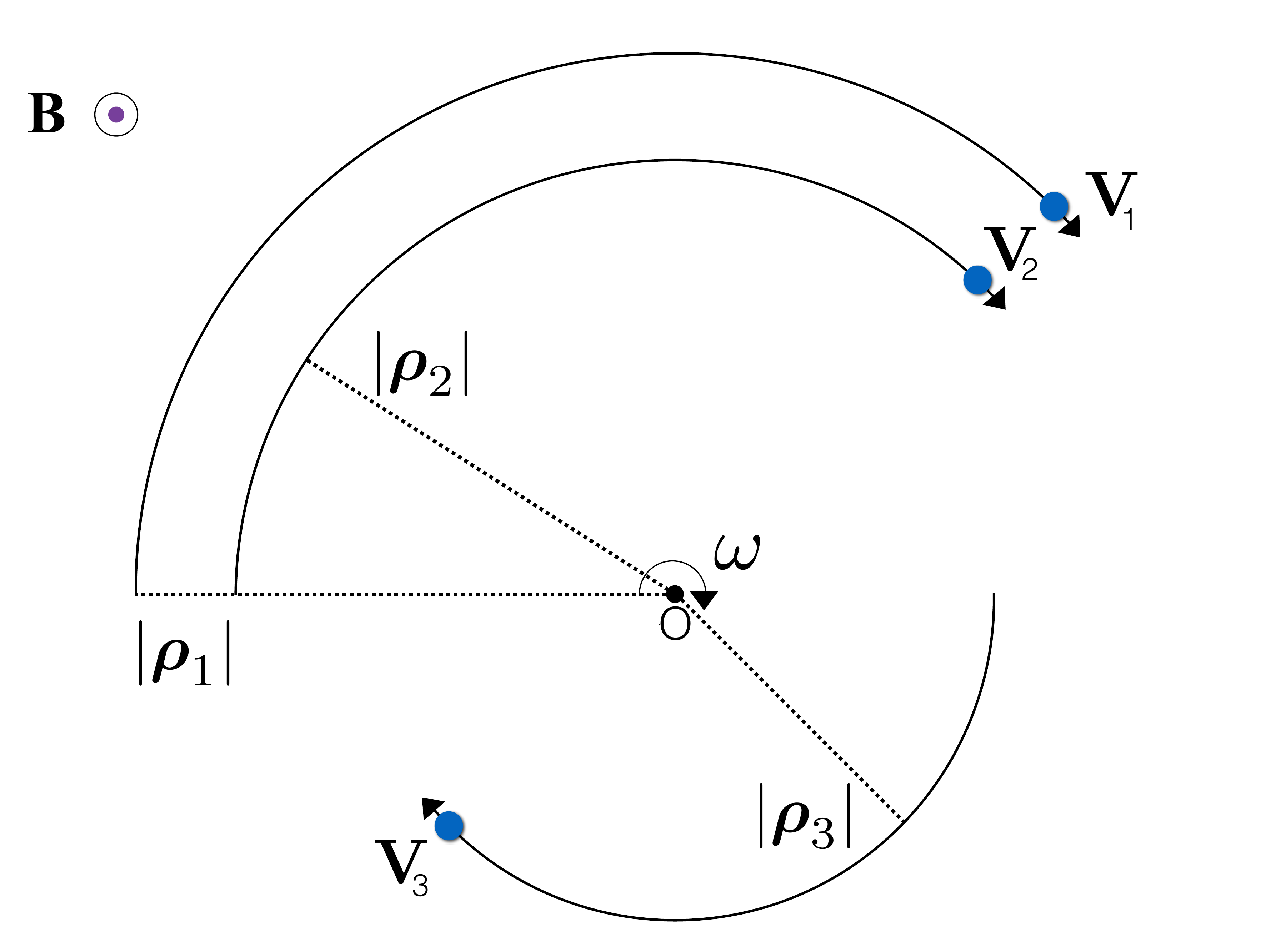}
\caption{\label{Config3}Configuration III. Two charges move in phase (counter)clockwise on two concentric circles with a relative phase $\pi$ with respect to the other charge. }
\end{figure}
\end{center}

The Configuration III corresponds to three particles rotating clockwise (or counterclockwise) with the same frequency $\omega$, in which two of them move in phase and the third one is shifted with a phase $\pi$. In this trajectory the not-in-phase particle can be located: a) along the inner radius, b) at the outer radius or c) in the central radius. The Configuration III for the case a) is presented in Fig. \ref{Config3}.

For the case a), the explicit form of the trajectories is given by

\begin{equation}
\begin{aligned}
& {\bo}_1(t) \ = \ \frac{{\rm v}_1}{\omega}\,(\,\cos \omega t  ,\, - \sin \omega t )  \ ,
\\ & {\bo}_2(t) \ = \ \frac{{\rm v}_2}{\omega}\,(\,\cos \omega t ,\, - \sin \omega t)  \ ,
\\ & {\bo}_3(t) \ = \ -\frac{{\rm v}_3}{\omega}\,(\,\cos \omega t ,\, -\sin \omega t ) \ .
\label{tC3}
\end{aligned}
\end{equation}

For convenience we assume the following conditions

\[
\rm{v}_1>\rm{v}_2>\rm{v}_3>0 \ ; \qquad \omega >0 \ .
\]

The concrete value of these parameters, as in the previous cases, is determined by the equations of motion (\ref{NEsep}).  The relative distances between the particles take the form

\begin{equation}
|{\bo}_1-{\bo}_2|\,=\,\frac{\rm{v}_1-\rm{v}_2}{\omega}\ , \qquad
|{\bo}_1-{\bo}_3|\,=\,\frac{\rm{v}_1+\rm{v}_3}{\omega}\ , \qquad
|{\bo}_2-{\bo}_3|\,=\,\frac{\rm{v}_2+\rm{v}_3}{\omega}\ .
\end{equation}

For the circular trajectories (\ref{tC3}), the Newton equations (\ref{NEsep}) lead to the following system of algebraic equations

\begin{equation}
\label{eqD21}
B\,e_1\,{\rm v}_1\ - \ m_1\,{\rm v}_1\,\omega \ - \  e_1\,\omega^2\,\bigg(\frac{e_2}{({\rm v}_1-{\rm v}_2)^2}+\frac{e_3}{({\rm v}_1+{\rm v}_3)^2}\bigg)\ = \ 0 \ ,
\end{equation}
\begin{equation}
\label{eqD22}
B\,e_2\,{\rm v}_2\ -  \ m_2\,{\rm v}_2\,\omega \  - \ e_2\,\omega^2\,\bigg(\frac{e_3}{({\rm v}_2+{\rm v}_3)^2}-\frac{e_1}{({\rm v}_1-{\rm v}_2)^2}\bigg)\ =\ 0 \ ,
\end{equation}
\begin{equation}
\label{eqD23}
-B\,e_3\,{\rm v}_3\  +   \  m_3\,{\rm v}_3\,\omega \  + \ e_3\,\omega^2\,\bigg(\frac{e_1}{({\rm v}_1+{\rm v}_3)^2}+\frac{e_2}{({\rm v}_2+{\rm
v}_3)^2}\bigg)\ =\ 0 \ .
\end{equation}

By adding the three equations (\ref{eqD21})-(\ref{eqD23}) we arrive to the relation between the magnetic field $B$ and the frequency $\omega$

\begin{equation}
\label{relD2}
\omega \ = \  \bigg(\frac{e_1 \,{\rm v}_1+e_2 \,{\rm v}_2-e_3\, {\rm v}_3  }{m_1 \,{\rm v}_1+m_2 \,{\rm v}_2-m_3 \,{\rm v}_3 }\bigg)\,B \ .
\end{equation}

The relation (\ref{relD2}) is a necessary (not sufficient) compatibility condition of the Eqs. (\ref{eqD21})-(\ref{eqD23}). Substituting (\ref{relD2}) into the equation (\ref{eqD22}) and solving for $B$ we obtain

\begin{equation}
\label{BD2}
B_{III} \ = - \frac{\left({\rm v}_1-{\rm v}_2\right){}^2 {\rm v}_2 \left({\rm v}_2+{\rm v}_3\right){}^2 \left(m_1 {\rm v}_1+m_2 {\rm v}_2-m_3 {\rm v}_3\right) \left(e_2 \left(m_1 {\rm v}_1-m_3 {\rm v}_3\right)+m_2 \left(e_3 {\rm v}_3-e_1 {\rm v}_1\right)\right)}{e_2 \left(e_1 \left({\rm v}_2+{\rm v}_3\right){}^2-e_3 \left({\rm v}_1-{\rm v}_2\right){}^2\right) \left(e_1 {\rm v}_1+e_2 {\rm v}_2-e_3 {\rm v}_3\right){}^2} \ .
\end{equation}

Notice that by replacing ${\rm v}_3\rightarrow- {\rm v}_3$ in $B_{III}$ we obtain $-B_{II}$ (see (\ref{BC2})).

Finally, putting Eqs. (\ref{relD2})-(\ref{BD2}) into the Eq. (\ref{eqD23}) we also get a sixth order polynomial equation in the variables ${\rm v}_1,\,{\rm v}_2,\,{\rm v}_3$ which coincides with (\ref{Pol2}) when ${\rm v}_3\rightarrow- {\rm v}_3$,

\begin{equation}
\label{Pol3}
\sum_{i+j+k=6}^{} a_{ijk}\,{\rm v}_1^i\,{\rm v}_2^j\,{(-{\rm v}_3)}^k \ = \ 0 \ .
\end{equation}

Therefore, its solutions can be obtained from those of (\ref{Pol2}). The system is \emph{particularly superintegrable}, and the corresponding integrals and \emph{particular} constants of motion are given by (\ref{IntMCII}) with the substitution $B_{II}\rightarrow B_{III}$ and ${\rm v}_3 \rightarrow - {\rm v}_3$.

It is worth to note that for particles with the special charge to mass ratio

\begin{equation}
\label{Wf2}
 \frac{e_1}{m_1} \ - \  \frac{(e_2 \,{\rm v}_2-e_3 \, {\rm v}_3)}{\left(m_2 \,{\rm v}_2-m_3\, {\rm v}_3\right)}  \ = \ 0 \ , \quad\quad\quad \bigg(\frac{e_1}{m_1}\neq \frac{e_3}{m_3}\bigg) \ ,
\end{equation}

the sixth order polynomial (\ref{Pol3}) becomes a second order polynomial in the ${\rm v}_1$ variable. The two independent solutions are given by the Eqs. (\ref{v1Wf}) with the substitution ${\rm v}_3\rightarrow-{\rm v}_3$. There exist special values of velocities and masses for which the neutral system $Q=0$ occurs in this Configuration III. The corresponding magnetic field $B$ is given by (\ref{BD2}).

For the cases in which the not-in-phase particle (see Fig. \ref{Config3}) is located either at the outer radius or in the central radius the corresponding special trajectories can be obtained straightforwardly.

\vspace{0.2cm}

\section{$N$-BODY CASE: SPECIAL TRAJECTORIES}
\label{STN}

Now, we proceed to study the case of $n\geq2$ Coulomb charges on the plane in a constant perpendicular magnetic field. The Hamiltonian is of the form \begin{equation}
\begin{aligned}
{\cal H}  \ = & \ \sum_{i=1}^{n}\frac{{({\bf p}_i-e_i\,{\bf A}_{{\bo}_{i}})}^2}{2\,m_i} \ + \ \sum_{i,j=1,2,...,n\,;\,j>i}^{} \frac{e_i\,e_j}{|\bo_i-\bo_j|}  \ ,
\end{aligned}
\label{Hindncase}
\end{equation}
where ${\bo}_{i}$ is the position vector of particle $i$, ${\bf p}_i$ is the associated canonical momentum and $\mathbf A_{\bf r}=\frac{1}{2}\ (\mathbf B\times \bf r)$. The total Pseudomomentum
\begin{equation}
{\bf K} \  \equiv \ (K_x,\,K_y) \ = \  \sum_{i=1}^{n}\boldsymbol k_i\ =\ \sum_{i=1}^{n} (\mathbf {p}_i\ +\ e_i\,\mathbf A_{{\bo}_{i}})\ ,
\label{pseudomomentumINDncase}
\end{equation}
is a constant of motion \cite{Avron}, the Poisson bracket $\{ {\bf K},\,{\cal {H}} \}=0$ vanishes. The total \emph{canonical} momentum ${\bf L}^{total}_z$
\begin{equation}
{\bf L}^{total}_z \     \equiv \ \sum_{i=1}^{n} \boldsymbol \ell_{z_i}  \  = \   \sum_{i=1}^{n}{\bo}_{i} \times {\bf p}_i \ ,
\label{LzTncase}
\end{equation}
is also conserved, $\{ {\bf L}^{total}_z, {\cal {H}} \}=0$. Hence, the problem is characterized at least by three conserved quantities (integrals) $K_{x,y},\ L_z^{total}$. The dimension of the configuration space is $2\,n$. The problem (\ref{Hindncase}) is not integrable, the number of integrals (including the Hamiltonian) is much less than the dimension of the configuration space. The integrals $K_{x,y},\ L_z^{total}$ are not in involution, they obey the commutation relations (\ref{AlgebraInt}) with $Q=e_1+e_2+\dots+e_n$.
From (\ref{Hindncase}) we obtain the Newton equations
\begin{equation}
\begin{aligned}
&m_i\,\ddot{\bo}_i\ =\ e_i\,\dot{\bo}_i\times \mathbf{B}\ + \ \sum_{j=1,2,...,n\,;\,j\neq i}^{} \frac{e_i\,e_j}{|\bo_i-\bo_j|^2}(\bo_i-\bo_j)\ , \qquad i=1,2,\ldots, n \ .
\end{aligned}
\label{NEsepncase}
\end{equation}
From (\ref{NEsepncase}) it follows that
\begin{equation}
\begin{aligned}
& \sum_{i=1}^{n}\bigg[\frac{m_i}{e_i}\,\ddot{\bo}_i \ - \  \dot{\bo}_i\times \mathbf{B}\bigg] \ = \ 0 \ .
\end{aligned}
\label{SCM}
\end{equation}

Therefore, from (\ref{SCM}) we arrive to the following interesting result: for the Hamiltonian ${\cal H}$ (\ref{Hindncase}) with particles of the same charge-to-mass ratio (equal Larmor frequencies)
\begin{equation}\label{CECM}
\frac{e_i}{m_i} \ = \ \alpha \ , \qquad i=1,2,\ldots,n \ ,
\end{equation}
exact separation of the CM occurs. Its motion is described by the equation $M\ddot{\bf R}\, =\, Q\,\dot{\bf R}\times \mathbf{B}$ and possesses the same Larmor frequency equal to $\alpha\,B$. For the physically important $n$-electron system, the condition (\ref{CECM}) is realized. This is the analog to the well known separation of variables for the free field case $B=0$.

\subsection{Special trajectories}

\begin{center}
\begin{figure}[h]
\hspace{1.5cm} \includegraphics[scale=0.2]{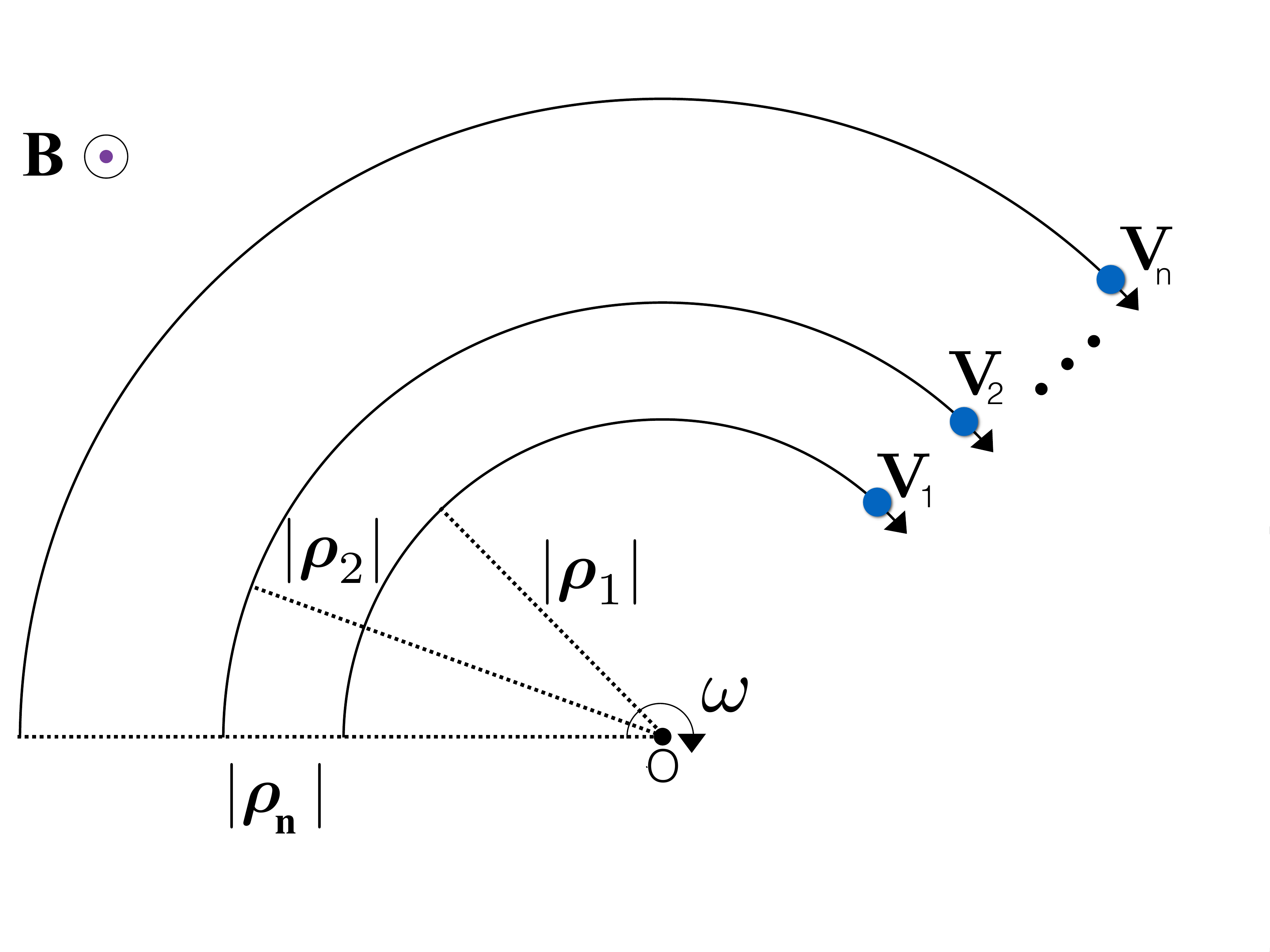}
\caption{\label{Config2n}Configuration II: special trajectories where the $n$ particles rotate clockwise (or counterclockwise) in phase with the angular frequency $\omega$.}
\end{figure}
\end{center}

The generalization of the Configuration II to the case of $n$ particles corresponds to the situation in which all the charges rotate clockwise (or counterclockwise) in phase with the same angular frequency $\omega$, (see Fig.(\ref{Config2n})). These special circular trajectories read

\begin{equation}
\begin{aligned}
& {\bo}_i(t) \ = \ \frac{{\rm v}_i}{\omega}\,(\,\cos \omega t  ,\, - \sin \omega t ) \quad , \qquad  i=1,2,\ldots,n \ ,
\label{tC2n}
\end{aligned}
\end{equation}

where without losing generality, we assume
\[
\rm{v}_i>\rm{v}_j>0\ , \quad (i>j) \quad ; \qquad \omega >0 \ .
\]
The concrete value of these quantities are determined by the equations of motion (\ref{NEsepncase}). The constant relative distances between the particles are given by
\begin{equation}
|{\bo}_i-{\bo}_j|=\frac{\rm{v}_i-\rm{v}_j}{\omega}\ , \,\quad i,j=1,2,\ldots, n \ ;\quad i>j \ ,
\end{equation}
all remain unchanged during time evolution.

Putting (\ref{tC2n}) into the Newton equations (\ref{NEsepncase}) we arrive to the system of $n$ coupled algebraic equations
\begin{equation}
\label{eqC21ncase}
B\,e_i\,{\rm v}_i\ - \ m_i\,{\rm v}_i\,\omega \  +  \ e_i\, \omega^2\,\sum_{j=1,2,...,n \,;\, i\neq j}^{}\frac{e_j\,s}{({\rm v}_i-{\rm v}_j)^2}\ = \ 0 \quad , \quad i=1,2,\ldots,n \ ,
\end{equation}
where $s=1$ for $j>i$ and $s=-1$ for $j<i$.

From (\ref{eqC21ncase}) it follows that for particles with equal charge to mass ratio
\[
\frac{e_i}{m_i} \ \equiv \ \alpha \quad , \qquad i=1,2,\ldots n \ ,
\]
with $\alpha \neq 0$ a real constant, the Configuration II does not occur. Therefore, $n$ electrons cannot perform this type of configuration.

\vspace{0.2cm}

Now, the sum of the $n$ equations (\ref{eqC21ncase}) gives the following relation between the magnetic field $B$ and the frequency $\omega$

\begin{equation}
\label{relC2ncase}
\omega \ = \  \bigg(\frac{e_1 \,{\rm v}_1+e_2 \,{\rm v}_2+\ldots + e_n\, {\rm v}_n  }{m_1 \,{\rm v}_1+m_2 \,{\rm v}_2+ \ldots + m_n \,{\rm v}_n }\bigg)\,B \ .
\end{equation}

The above relation (\ref{relC2ncase}) is a necessary (not sufficient) condition for the existence of special trajectories in the Configuration II.  Substituting (\ref{relC2ncase}) into the equation of motion in (\ref{eqC21ncase}) with $i=2$ and then solving for $B\equiv B_{II}$ we obtain

\begin{equation}
\label{BD2n}
B_{II} \ = \  \frac{{\rm v}_2\,(m_2\,{\kappa}-e_2)}{e_2\,{{\kappa}}^2}{\bigg[\sum_{j=3}^{n}\frac{e_j}{({\rm v}_2-{\rm v}_j)^2}-\frac{e_1}{({\rm v}_2-{\rm v}_1)^2}\bigg] }^{-1} \ ,
\end{equation}

${\kappa}=\frac{e_1 \,{\rm v}_1+e_2 \,{\rm v}_2+\ldots + e_n\, {\rm v}_n  }{m_1 \,{\rm v}_1+m_2 \,{\rm v}_2+ \ldots + m_n \,{\rm v}_n }$.

In general, substituting (\ref{BD2n}) into (\ref{eqC21ncase}) we get a system of $(n-2)$ multivariate polynomial coupled equations of order $(4n-6)$ for the parameters ${\rm v}_1,\,{\rm v}_2,\ldots,{\rm v}_n$

\begin{equation}
\label{Pol3}
\sum_{g_1+g_2+\ldots+g_n=4n-6}^{} a^{(l)}_{g_1\,g_2\ldots g_n}\,{\rm v}_1^{g_1}\,{\rm v}_2^{g_2}\,\ldots\,{\rm v}_n^{g_n} \ = \ 0 \ , \qquad l=1,2,\ldots,n-2 \ ,
\end{equation}

where the coefficients $a^{l}_{ijk}=a^{l}_{ijk}(e_1,e_2,\ldots,e_n,m_1,m_2,\ldots,m_n)$. Therefore, formally there exist a two-parametric family of initial conditions for which special trajectories appear.

As for the integrals and \emph{particular} constants of motion, in addition to the three integrals $K_{x,y},\ L_z^{total}$ (\ref{pseudomomentumINDncase})-(\ref{LzTncase}) there exist $2\,(n-1)$ \emph{particular} constants of motion, namely $(n-1)$ individual angular momenta $ \boldsymbol \ell_{z_i}$ and $(n-1)$ kinetic terms $\frac{1}{2}m_i\,{\rm v}_i^2$. The system is \emph{particularly superintegrable}.

\section{CONCLUSIONS}
\label{Concl}

A classification of systems with three charged particles on the plane placed in a perpendicular constant magnetic field $B$, which admit special trajectories was presented. In general, these trajectories describe concentric circles of finite radii. Their main characteristic is that relative distances between particles remain unchanged under the time evolution. Similar to the two body case, it
corresponds to the existence of \emph{particular} constants of motion.

These special periodic trajectories are characterized by seven conserved quantities. The three integrals $K_x,\,K_y,\,L_z^{total}$ which are conserved for any trajectory (arbitrary initial conditions) and four \emph{particular} constants that emerge only for certain values of initial data. Hence, these trajectories are \emph{particularly superintegrable}. The complete classification of such initial data was presented in detail.

There are three important physical systems admitting special trajectories:

\begin{itemize}
  \item $Q\,=\,0$ (neutral system), the special trajectories of all Configurations I, II and III appear.
  \item $\frac{e_1}{m_1}=\frac{e_2}{m_2}=\frac{e_3}{m_3}=\alpha$\, (particles with equal Larmor radius): the special trajectories of Configuration I occur only.
  \item $e_1\,=-2\,e\,;e_2=e_3=e$ (Helium-like system), all Configurations I, II and III appear.
\end{itemize}

Along these lines, results for the $n$-body problem in a constant magnetic field were presented as well. The separation of the center of mass for particles with the same charge to mass ratio (an $n$-electron system) and, in general, the existence of a non-trivial two-parametric family of special periodic trajectories were indicated.

The issue about the stability of the special trajectories was not addressed in the present work since it is not relevant for the quantum case which is the main goal we are interested in. After the standard quantization of the Hamiltonian $\cal H$ (\ref{Hind}), the Pseudomomentum  (\ref{pseudomomentumIND}), the angular momentum (\ref{LzT}), the \emph{particular} constants of motion, i.e. upon replacing the momenta by the corresponding differential operators, one can ask whether there exist eigenstates which are common for $\cal H$ and one of the \emph{particular} constants. Such common eigenfunctions may exist, as in the two-body problem \cite{AMT}, for certain discrete values of the magnetic field $B$ and even for systems that classically are known to be chaotic like the neutral system. In this paper we have identified all the physical systems, the integrals and \emph{particular} constants of motion that may lead to exact solutions of the quantum three body Coulomb problem in a magnetic field. This construction would imply a certain particular integrability: the commutator (the Lie bracket) of the Hamiltonian and an operator vanishes on a subspace of the Hilbert space \cite{Turbiner:2013p}.

\section{ACKNOWLEDGMENTS}

M.A.E.R. is grateful to ICN UNAM, Mexico for the kind hospitality during his visit, where a part of the research was done as well as to the CRM, Montreal, where it was completed. He was supported in part by DGAPA grant {\bf IN108815} (Mexico) and, in general, by a fellowship awarded by the Laboratory of Mathematical Physics of the CRM for postdoctoral research. He is deeply grateful to A. Turbiner for the proposal of this problem and for useful discussions and important remarks during the early stage of the work. C. A. E. was supported by a CONACyT postdoctoral grant No. 234745.

\section{Appendix}
\label{App}

The multivariate polynomial equation of sixth order in the variables ${\rm v}_1,\,{\rm v}_2,\,{\rm v}_3$ solutions of which give rise to the Configuration II (Fig. \ref{Config2}) is given by

{
\footnotesize

\begin{equation}\label{}
\begin{aligned}
&\mathbb{P}_6 \ =\  e_2 e_3 \left(e_2 m_1-e_1 m_2\right)\, {\rm v}_2 {\rm v}_1^5 \ + \ e_2 e_3 \left(e_3 m_1-e_1 m_3\right) {\rm v}_3 {\rm v}_1^5 \ + \ 2 e_2 e_3 \left(e_1 m_2-e_2 m_1\right)\, {\rm v}_2^2 {\rm v}_1^4 \ + \ 2 e_2 e_3 \left(e_1 m_3-e_3 m_1\right) {\rm v}_3^2 {\rm v}_1^4
\\ &
+ \ 2 e_2 e_3 \left(e_1 \left(m_2+m_3\right)-e_2 m_1-e_3 m_1\right)\, {\rm v}_2 {\rm v}_3 {\rm v}_1^4 \ - \ \left(e_1+e_2\right) e_3 \left(e_1 m_2-e_2 m_1\right) {\rm v}_2^3 {\rm v}_1^3 \ + \ e_2 \left(e_1-e_3\right) \left(e_1 m_3-e_3 m_1\right)\, {\rm v}_3^3 {\rm v}_1^3
\\ &
+\ \left(  e_2 e_3 \left(3 m_1-m_2-4 m_3\right) e_1 -\left(e_3 m_2+2 e_2 m_3\right) e_1^2  +  e_2 e_3 \left(e_2+4 e_3\right) m_1\right) \, {\rm v}_2 {\rm v}_3^2 {\rm v}_1^3
\\ &
+\ \left(\left(2 e_3 m_2+e_2 m_3\right) e_1^2-e_2 e_3 \left(3 m_1+4 m_2+m_3\right) e_1+e_2 e_3 \left(4 e_2+e_3\right) m_1\right)\, {\rm v}_2^2 {\rm v}_3 {\rm v}_1^3
\\ &
 +\ 2 e_1 e_3 \left(e_1 m_2-e_2 m_1\right)\, {\rm v}_2^4 {\rm v}_1^2 \ + \ 2 e_1 e_2 \left(e_3 m_1-e_1 m_3\right) \, {\rm v}_3^4 {\rm v}_1^2
\\ &
 + \ \left[4 e_2 m_3 e_1^2+\left(m_3 e_2^2-e_3 \left(4 m_1+m_2-3 m_3\right) e_2-e_3^2 m_2\right) e_1-2 e_2 e_3^2 m_1\right] \,{\rm v}_2 {\rm v}_3^3 {\rm v}_1^2
\\ &
 -\ 2 \left[\left(e_3 m_1+e_1 m_3\right) e_2^2+\left(m_3 e_1^2-2 e_3 m_2 e_1+e_3^2 m_1\right) e_2-e_1 e_3 \left(e_1+e_3\right) m_2\right] \, {\rm v}_2^2 {\rm v}_3^2 {\rm v}_1^2
\\ &
 +\ \left[\left(m_3 e_2^2+e_3 \left(4 m_1+m_2+m_3\right) e_2-e_3^2 m_2\right) e_1-4 e_3 m_2 e_1^2-2 e_2^2 e_3 m_1\right]\,  {\rm v}_2^3 {\rm v}_3 {\rm v}_1^2
\\ &
+\ e_1 e_3 \left(e_2 m_1-e_1 m_2\right)\, {\rm v}_2^5 {\rm v}_1 \ + \ e_1 e_2 \left(e_1 m_3-e_3 m_1\right) {\rm v}_3^5 {\rm v}_1 \ + \ 2 e_1 e_2 \left(e_3 \left(m_1+m_2\right)-\left(e_1+e_2\right) m_3\right)\, {\rm v}_2 {\rm v}_3^4 {\rm v}_1
\\ &
 + \ \left[e_2 m_3 e_1^2+\left(4 m_3 e_2^2-e_3 \left(m_1+4 m_2+3 m_3\right) e_2+2 e_3^2 m_2\right) e_1+e_2 e_3^2 m_1\right]\, {\rm v}_2^2 {\rm v}_3^3 {\rm v}_1
\\ &
 + \ \left[\left(e_3 \left(m_1+m_2+4 m_3\right) e_2-2 m_3 e_2^2-4 e_3^2 m_2\right) e_1-e_3 m_2 e_1^2+e_2^2 e_3 m_1\right]\, {\rm v}_2^3 {\rm v}_3^2 {\rm v}_1
\\ &
 + \ 2 e_1 e_3 \left(\left(e_1+e_3\right) m_2-e_2 \left(m_1+m_3\right)\right)\, {\rm v}_2^4 {\rm v}_3 {\rm v}_1\ + \ e_1 e_2 \left(e_2 m_3-e_3 m_2\right)\, {\rm v}_2 {\rm v}_3^5 \ + \ 2 e_1 e_2 \left(e_3 m_2-e_2 m_3\right)\, {\rm v}_2^2 {\rm v}_3^4
\\ &
 + \ e_1 \left(e_2+e_3\right) \left(e_2 m_3-e_3 m_2\right)\, {\rm v}_2^3 {\rm v}_3^3
 \ + \ 2 e_1 e_3 \left(e_3 m_2-e_2 m_3\right)\, {\rm v}_2^4 {\rm v}_3^2 \ + \ e_1 e_3 \left(e_2 m_3-e_3 m_2\right)\, {\rm v}_2^5 {\rm v}_3 \ = \ 0 \ .
\end{aligned}
\end{equation}

}

\end{document}